\documentclass[aps,amsmath,amssymb,reprint,twocolumn,prl,superscriptaddress,notitlepage,showpacs,tightenlines]{revtex4-2}

\usepackage{exscale}
\usepackage{relsize}
\usepackage{graphicx}
\usepackage{dcolumn}
\usepackage{bm}
\usepackage{textcomp}
\usepackage[utf8]{inputenc}
\usepackage[T1]{fontenc}
\usepackage{mathptmx}
\usepackage{siunitx}
\usepackage{array}
\usepackage{color}
\usepackage{textcomp}
\usepackage{amsmath}
\usepackage{float}
\usepackage{verbatim}
\usepackage[colorlinks,citecolor=blue,linkcolor=red,hyperindex,CJKbookmarks]{hyperref}
\usepackage{natbib}

\allowdisplaybreaks[1]

\begin{document}
\title{Electromagnetic Selection Rules and Coherent Manipulation of Quantum Skyrmion via Surface Acoustic Wave Phonons}

\author{Geng Li}
\thanks{These authors contributed equally to this work.}
\affiliation{School of Integrated Circuits, Tsinghua University, Beijing 100084, China}
	
\author{Yu-Yuan Chen}
\thanks{These authors contributed equally to this work.}
\affiliation{School of Integrated Circuits, Tsinghua University, Beijing 100084, China}
	
\author{Yu-xi Liu}
\email{yuxiliu@mail.tsinghua.edu.cn}
\affiliation{School of Integrated Circuits, Tsinghua University, Beijing 100084, China}
\affiliation{Frontier Science Center for Quantum Information, Beijing 100084, China}
\date{\today}

\begin{abstract}
Skyrmions are competitive candidates for information-storage units and have great prospect in quantum information processing. We here study coherent control of a quantum skyrmion via surface acoustic wave (SAW) phonons in a piezoelectric SAW cavity. Exact diagonalization of a finite Dzyaloshinskii-Moriya cluster reveals the eigenstates of quantum skyrmion with robust scalar chirality. The underlying lattice-spin symmetry imposes polarization-dependent selection rules for the electromagnetic transitions between eigenstates states. Considering the small mode volume of the SAW phonon easy to reach strong coupling, we derive the interaction Hamiltonian between the quantum skyrmion as a qubit and a single-mode quantized SAW via the electric field induced by piezoelectric effect, and find that the coupling strength at the single-phonon level increases linearly with skyrmion radius. This enables enhanced spin-acoustic coupling for larger skyrmionic textures and for low magnetic dissipation. Our results establish few-spin quantum skyrmions as compact building blocks for hybrid quantum information processing and suggest a route toward more densely integrated on-chip quantum devices.
\end{abstract}
\maketitle
	
\textcolor{blue}{\it Introduction.---}
Magnetic skyrmions are nanoscale noncollinear spin textures characterized by chiral winding and nontrivial topology~\cite{SkyrmionFirstTheory,SkyrmionReviewDevices1,SkyrmionReviewDevices2,SkyrmionReview1}. Their topological stability, small size, and controllable dynamics make them attractive candidates for information carriers at the classical level~\cite{SkyrmionReviewDevices2,SkyrmionReviewDevices3,SkyrmionReview1}.
Recently, growing attention has been devoted to the quantum nature of skyrmions~\cite{SkyrmionReview2}. For example,  the skyrmion qubits have been proposed for quantum information processing, and correspond to quantization of the continuous collective coordinate of the semiclassical skyrmions~\cite{SkyrmionQubitFirstTheory,SkyrmionQubitFirstTheoryHelicity,SkyrmionQubitFirstReview,SkyrmionQubitQuantumComputing}. Differently, in the few-spin limit, a quantum skyrmion is regarded as a fully quantum realization of a skyrmion: a discrete, strongly correlated many-body state in frustrated or Dzyaloshinskii-Moriya-interaction magnets, which can exhibit field-stabilized scalar chirality~\cite{QuantumSkyrmionFrustratedMagnet,QuantumSkyrmionStability,QuantumSkyrmionPatterns,Frustrated3,
QuantumSkyrmionControlledCreation,SkyrmionicQubitsDMI,HallerDiazBelzigSchmidt2024}.~Radio-frequency fields, including microwave electromagnetic fields and surface acoustic waves (SAWs), have emerged as effective tools for manipulating classical magnetic skyrmions~\cite{SkyrmionModulationField,SkyrmionMotionControl1,SkyrmionMotionControl2,SkyrmionMotionControl3}.

SAWs are mechanical waves and propagate along the surface of an elastic material. In a piezoelectric material, the mechanical vibration induces electric fields via  the piezoelectric effect, enabling SAWs to couple to solid-state quantum systems through strain and electric  interactions~\cite{SAWBackgroundBook1,SAWBackgroundBook2,QuantumAcousticsArticle2,QuantumAcousticsSAWSuperconductingCoupling1,QuantumAcousticsTransducer1,Chen2025PiezomagneticSAW,QuantumAcousticsSAWSuperconductingCoupling2}.
SAWs propagate about five orders of magnitude slowly than microwave electromagnetic fields  at a given frequency, yielding wavelengths that are correspondingly much shorter. This allows compact on-chip confinement and spatially localized quantum control~\cite{SAWBackgroundBook1,SAWBackgroundBook2,QuantumAcousticsArticle2}.
These properties make SAWs a versatile interface for manipulating microscale quantum systems~\cite{QuantumAcousticsTransducer1,QuantumAcousticsSAWSuperconductingCoupling1,QuantumAcousticsArticle2}. Over the past decade, the quantized SAWs coupled to superconducting qubits have shown potential in quantum control of acoustic modes or qubit states, phonon-mediated quantum-state transfer, and remote entanglement~\cite{satzinger2018quantum,Oleg,SAWEntanglement1,SAWEntanglement2,QuantumAcousticsMultimodeSAW,SAWCavityMultimodeTunableCoupling}.
Moreover, SAWs provide a promising on-chip data bus for interfacing superconducting qubits~\cite{SuperconductingCircuit1,SuperconductingCircuit2}~with magnetic systems~\cite{Chen2025PiezomagneticSAW,HybridSAWSkyrmion}. This enables hybrid architectures that combine controllable superconducting qubits with long-lived magnetic quantum systems~\cite{HybridSAWSkyrmion,SuperconductingHybrid,SuperconductingHybridChip,PhysRevLett.132.193601}.

We here develop a theory for coherent manipulation on a quantum skyrmion by SAW phonons in a piezoelectric SAW cavity. Previous studies have shown the existence, stability, ground-state properties, and optical signature of quantum skyrmions~\cite{QuantumSkyrmionFrustratedMagnet,Frustrated3,QuantumSkyrmionControlledCreation,QuantumSkyrmionStability,QuantumSkyrmionPatterns,SharmaPsaroudaki2026}. However, the microscopic energy-level structure and electromagnetic transition selection rules between different eigenstates of quantum skyrmions remain to be established for coherent manipulation. By exactly diagonalizing quantum skyrmions, we show the low-energy skyrmionic states characterized by its eigenenergy spectrum, scalar chirality, and coherent-superposition.~The corresponding polarization-dependent selection rules for electromagnetic transitions are also studied. We find that the coupling strength between the quantum skyrmion and the quantized microwave field at single-photon level is much smaller than that between the quantum skyrmion and quantized SAW at single-phonon level.  Based on these, we further study the coupling of the quantum skyrmion as a qubit to the quantized SAW~\cite{SkyrmionQubitPhonon2024}. Different from collective-coordinate descriptions of semiclassical skyrmion qubits~\cite{SkyrmionQubitFirstTheory,SkyrmionQubitFirstTheoryHelicity}, our approach directly resolves the microscopic many-body states and provides a route to manipulate a quantum-skyrmion qubit at a single-phonon level.

\begin{figure}[t]
	\centering
	\includegraphics[width=1.0\columnwidth]{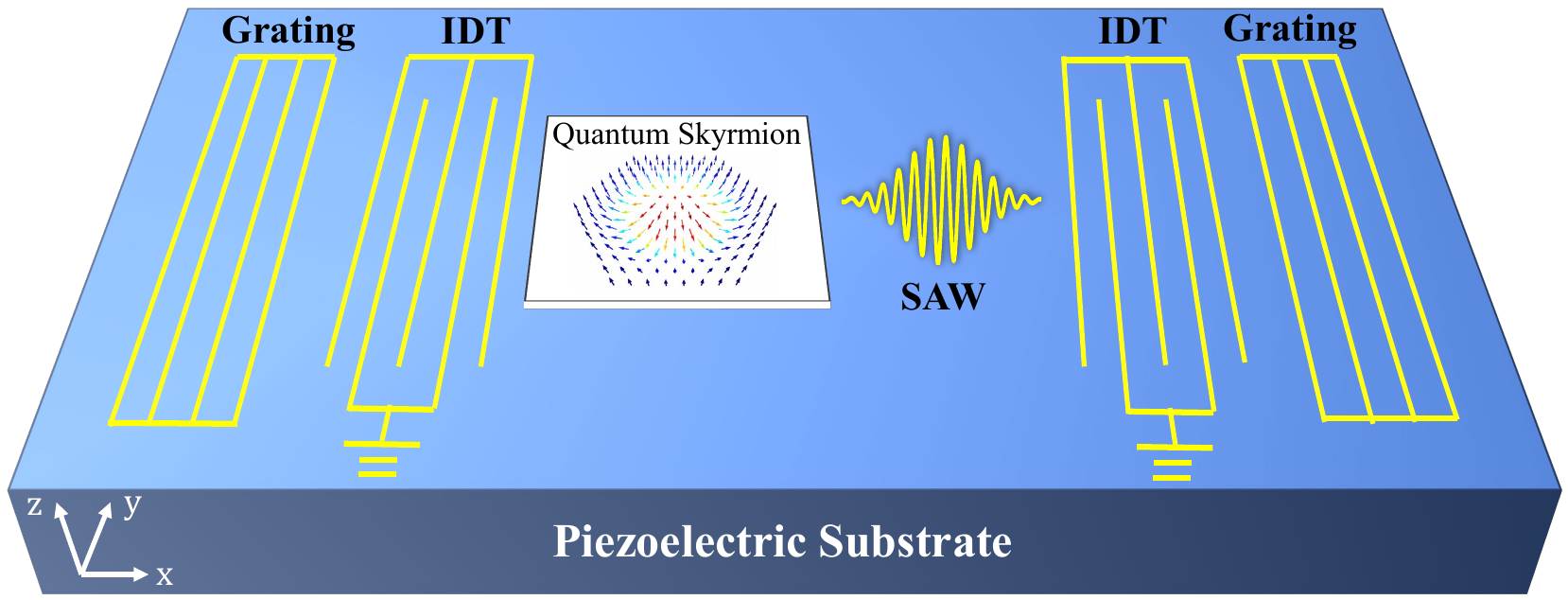}
	\caption{Schematic of the hybrid spin-acoustic system. A quantum skyrmion hosted in a magnetic nanostructure is integrated on a piezoelectric substrate. IDTs excite and detect SAWs, while Bragg gratings form a SAW cavity. The confined SAW phonon is coupled to the spin-induced electric polarization of the quantum skyrmion via the electric field produced through piezoelectric material.}
    \label{fig:systemschematic}
\end{figure}

\textcolor{blue}{\it Theoretical model and eigenenergy spectra.---}
As schematically shown in Fig.~\ref{fig:systemschematic}, we study an integrated spin-acoustic system  that a quantum skyrmion hosted in a magnetic nanostructure is coupled to a SAW cavity, fabricated on a piezoelectric substrate, via the electric field induced by either the classical or the quantized SAW through the piezoelectric effect~\cite{Gustafsson2014,Manenti2017,KNBMechanism}. Here, interdigital transducers (IDTs) are used to excite and detect SAWs in cavity, which is formed by two Bragg gratings~\cite{SAWCavity}. This device enables coherent phononic control of the quantum-skyrmion via the electric field.
	
We here focus on the quantum skyrmion model, described by the typical Dzyaloshinskii-Moriya-Interaction (DMI) Hamiltonian on a finite triangular lattice with $\hat H_{\mathrm{DMI}}=J\sum_{\langle i,j\rangle}\hat{\bm S}_i\!\cdot\!\hat{\bm S}_j+\sum_{\langle i,j\rangle}\bm D_{ij}\!\cdot\!\left(\hat{\bm S}_i\times \hat{\bm S}_j\right)+B\sum_i \hat S_i^z$, with the Pauli spin operators $\hat{\bm S}_i$ and $\hat S_i^z$. $\langle i,j\rangle$ denotes an unordered nearest-neighbor bond of the triangular lattice. $J$ is the exchange coupling~\cite{Heisenberg1928}. $\mathbf{D}_{ij}$ is the in-plane DMI vector perpendicular to the bond connecting sites $i$ and $j$~\cite{Dzyaloshinskii1958,Moriya1960}, thereby favoring N\'eel-type skyrmionic textures~\cite{SkyrmionReviewDevices2,QuantumSkyrmionFrustratedMagnet}. $B$ is an external magnetic field along $z$. This Hamiltonian describes a DMI-based model widely employed in exact-diagonalization studies of finite clusters and few-spin skyrmionic states~\cite{QuantumSkyrmionFrustratedMagnet,Frustrated3,
	QuantumSkyrmionStability,QuantumSkyrmionPatterns,QuantumSkyrmionControlledCreation,SkyrmionicQubitsDMI}.

For finite clusters, however, the skyrmionic states depend sensitively on the boundary condition. Periodic boundary conditions (PBC)~\cite{QuantumSkyrmionFrustratedMagnet,Frustrated3,QuantumSkyrmionStability,QuantumSkyrmionPatterns}~supports quantum-skyrmion phases, characterized by scalar chirality, spin correlations, and symmetry-resolved low-energy structure~\cite{QuantumSkyrmionFrustratedMagnet,QuantumSkyrmionStability,QuantumSkyrmionPatterns}. Differently, open boundary conditions (OBC)~\cite{SkyrmionicQubitsDMI} introduces important edge effects, leading to more classical-like skyrmionic textures~\cite{BoundaryTuningSkyrmion}. Most existing studies have adopted either PBC or OBC, however, neither of which faithfully captures a isolated skyrmion in a finite cluster. To interpolate between PBC and OBC, we therefore partition the bonds into internal bonds $C_{\mathrm{in}}$ and boundary-connecting bonds $C_{\mathrm{bd}}$, and introduce a phenomenological parameter $\lambda$ for continuously tuning the boundary-connecting bond strength. Thus, we obtain
\begin{align}
	H(\lambda)={}& \sum_{\langle i,j\rangle}
	w_{ij}(\lambda)
	\left[
	J\,\mathbf{S}_i\cdot\mathbf{S}_j
	+\mathbf{D}_{ij}\cdot
	\left(\mathbf{S}_i\times\mathbf{S}_j\right)
	\right]
	+B\sum_i S_i^z ,
	\label{eq:H_lambda}
\end{align}
where the bond weight takes $w_{ij}(\lambda)=1$ for $\langle i,j\rangle\in C_{\rm in}$ and $w_{ij}(\lambda)=\lambda$ for $\langle i,j\rangle\in C_{\rm bd}$, so that $\lambda$ simultaneously tunes the exchange and DMI interactions on the boundary-connecting bonds. Thus, $\lambda=0$ and $\lambda=1$ recover the OBC and PBC limits, respectively, while $0<\lambda<1$ describes intermediate boundary connectivity. This interpolation is built upon previous studies that boundary engineering and edge confinement effectively reshape skyrmionic states~\cite{QuantumSkyrmionControlledCreation,BoundaryTuningSkyrmion}. The corresponding eigenvalues $E_n(\lambda,B)$ and eigenstates $|\psi_n(\lambda,B)\rangle$ satisfy
\begin{equation}
	\hat H(\lambda)\,
	|\psi_n(\lambda,B)\rangle=
	E_n(\lambda,B)\,
	|\psi_n(\lambda,B)\rangle ,
	\label{eq:eigen_problem}
\end{equation}
for different $\lambda$ and $B$, allowing us to track the evolution of the energy spectrum across the OBC-PBC crossover. For concreteness, we take $D_{ij}\equiv D=1$ and $J=-0.5$ in the following calculations.

\begin{figure}[tb]
		\centering
		\includegraphics[width=\columnwidth]{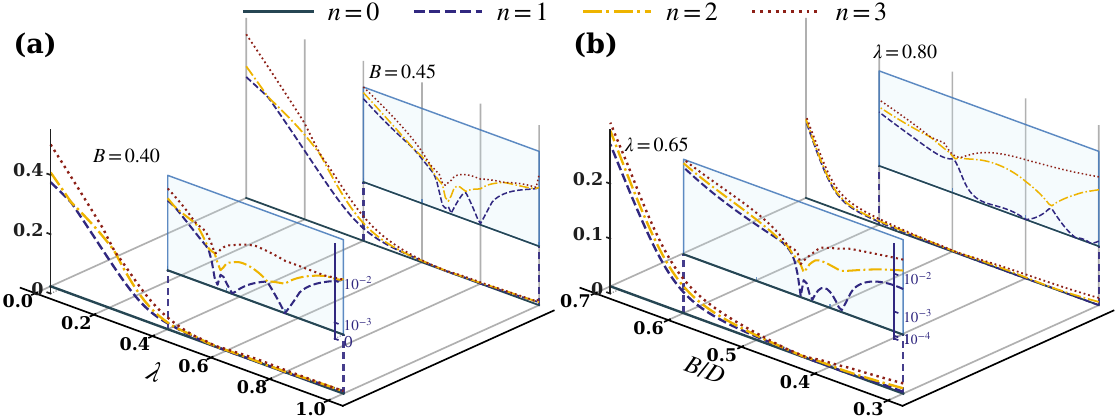}
		\caption{Normalized energy spectra \((E_n-E_0)/D\) of the model in  Eq.~(\ref{eq:H_lambda}). (a) Eigenvalues as functions of the boundary parameter \(\lambda\) for two magnetic fields, \(B=0.40\) and \(B=0.45\), shown as stacked spectra. (b) Eigenvalues as functions of the normalized magnetic field \(B/D\) for two boundary parameters, \(\lambda=0.65\) and \(\lambda=0.80\), shown as stacked spectra. The different colors and line styles identify the seven lowest eigenstates, \(n=0,\ldots,3\). The insets show the corresponding low-energy spectra on a logarithmic energy scale over $\lambda\in[0.4,1.0]$ in (a) and $B/D\in[0.3,0.6]$ in (b), respectively.}
		\label{fig:model_spectrum}
\end{figure}

Figure~\ref{fig:model_spectrum}(a) plots $E_n(\lambda,B)$ versus $\lambda$ for different magnetic fields $B$, respectively. It shows that the energy levels are strongly reshaped by the boundary condition, characterized by $\lambda$. At $\lambda=1$, the skyrmions with PBC recovers the higher rotational symmetry. This restoration of the symmetry appears as the multiplet structure near $\lambda=1$, including the threefold degeneracy in Fig.~\ref{fig:model_spectrum}(a), which is consistent with the symmetry-resolved quantum-skyrmion structure previously identified in periodic triangular lattice settings~\cite{QuantumSkyrmionFrustratedMagnet,Frustrated3}.  In Fig.~\ref{fig:model_spectrum}(b), we also show the variations of $E_n(\lambda,B)$ versus the tunable magnetic field $B$ at fixed boundary condition described by given $\lambda$. Thus, the parameter $\lambda$ provides a transparent theoretical interpolation between open and periodic geometries, whereas the magnetic field $B$ provides a realistic way to tune the energy spetra. In the ceratin parameter region where the lowest two energy levels are well isolated from higher energy levels, they can be selected as an effective two-level system for the skyrmion qubit  (see, supplemental materials~\cite{SupplementalMaterial}.)

\textcolor{blue}{\it Diagnostic of quantum skyrmions via scalar chirality.---}Scalar chirality is a standard diagnostic of noncoplanar many-spin correlations and a key indicator of quantum-skyrmion states in finite frustrated and DMI-based clusters~\cite{QuantumSkyrmionFrustratedMagnet,Frustrated3,QuantumSkyrmionControlledCreation,QuantumSkyrmionStability,QuantumSkyrmionPatterns}. The scalar chirality of $n$th eigenstate $|\psi_n(\lambda,B)\rangle$ is defined as
\begin{equation}
	Q_n(\lambda,B)=
	\frac{1}{8\pi}\langle \psi_n(\lambda,B)|\sum_{\langle ijk\rangle}\hat{\mathbf S}_i\cdot(\hat{\mathbf S}_j\times \hat{\mathbf S}_k)|\psi_n(\lambda,B)\rangle,
	\label{eq:quantum_topological_charge}
\end{equation}
where $\langle ijk\rangle$ runs over the elementary three-site triangular plaquettes, with $(i,j,k)$ ordered counterclockwise relative to a surface normal to $z$-direction. This orientation fixes the sign of $\mathbf{S}_i\cdot(\mathbf{S}_j\times\mathbf{S}_k)$ and ensures a consistent summation of the local scalar chiralities~\cite{QuantumSkyrmionFrustratedMagnet,Frustrated3}. Unlike the classical topological charge, $Q_n$ is not a mathematically exact topological invariant in a finite spin system~\cite{QuantumSkyrmionStability,QuantumSkyrmionPatterns}, because quantum fluctuations remain intrinsically present. Nevertheless, it provides a physical characteristic of the quantum skyrmions, especially used together with the corresponding energy spectra and the symmetry of eigenstates~\cite{QuantumSkyrmionFrustratedMagnet,QuantumSkyrmionStability,QuantumSkyrmionPatterns}. Here, $Q_n$ is evaluated directly for an individual many-body eigenstate, whereas the topological charge characterizes an effective real-space spin texture.

\begin{figure}[tb]
	\centering
	\includegraphics[width=\columnwidth]{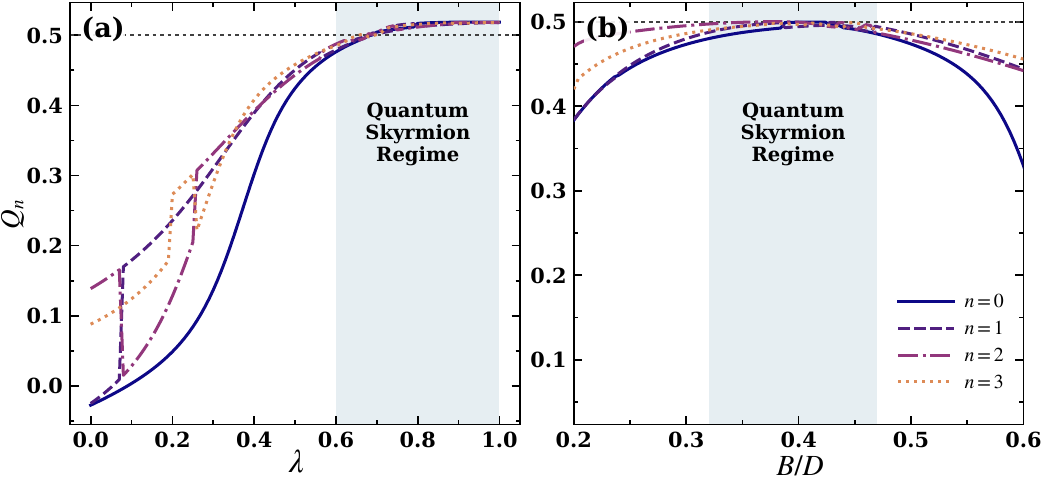}
	\caption{Quantum scalar chirality $Q_n$ of the low-lying eigenstates, defined by Eq.~(\ref{eq:quantum_topological_charge}). (a) $Q_n$ as a function of the boundary parameter $\lambda$ at fixed $B=0.45$. (b) $Q_n$ as a function of magnetic field $B$ at fixed $\lambda=0.65$. The horizontal dashed line marks $Q=0.5$, characteristic of the quantum-skyrmion regime in the triangular-lattice model with total number $N=19$ of spins. Colors and line styles identify the same eigenstate index $n$ in both panels.}
	\label{fig:topological_charge_quantum}
\end{figure}

Figure~\ref{fig:topological_charge_quantum}(a) shows $Q_n$ versus $\lambda$ at fixed magnetic field $B=0.45$ for several low-lying eigenstates. When the  boundary approaches the periodic limit $\lambda \simeq 0.6$-$1$, it is found that the ground-state scalar chirality remains close to $Q_0\simeq 0.5$, the characteristic value previously reported for the quantum-skyrmion regime of the $N=19$ triangular-lattice model under PBC~\cite{QuantumSkyrmionStability,QuantumSkyrmionPatterns}.~Thus, the skyrmionic character maintains even when the system is not exactly at perfect PBC. Figure~\ref{fig:topological_charge_quantum}(b) plots $Q_n$ versus $B$ at fixed boundary $\lambda=0.65$. Here again, scalar chirality remains near $0.5$ within a finite field window $B=0.32$-$0.47$, indicating that the low-energy skyrmionic states are robust to not only deviation from PBC but also moderate variation of the magnetic field. This behavior is consistent with earlier results, where the scalar chirality remains finite and nearly constant throughout the quantum-skyrmion phase, but changes rapidly upon entering the topologically trivial regime~\cite{QuantumSkyrmionFrustratedMagnet,QuantumSkyrmionStability,QuantumSkyrmionPatterns}.

Importantly, arbitrary coherent superpositions within the two lowest skyrmionic states retain a scalar chirality close to $0.5$ over the complete logical Bloch sphere, as verified in Ref.~\cite{SupplementalMaterial}. Thus, despite the coherent manipulation between different many-body eigenstates, the low-energy states remains a chirality-preserving logical subspace, consistent with previous studies of quantum-skyrmion dynamics~\cite{QuantumSkyrmionFrustratedMagnet,Frustrated3,QuantumSkyrmionStability}. This robustness makes the quantum skyrmion a suitable logical subspace for electromagnetic or phononic control: coherent evolution can modify the state without immediately destroying its skyrmionic character. A comparison with the classical texture-based diagnostic of a coherently reconstructed state is provided in Ref.~\cite{SupplementalMaterial}.

\textcolor{blue}{\it Electromagnetic control and transition selection rules.---}
The manipulation of the quantum skyrmions can be realized via either the electric~\cite{QuantumAcousticsTransducer1} or the magnetic fields~\cite{Chen2025PiezomagneticSAW}. We here focus the manipulation on quantum skyrmions via an electric field, described by the interaction Hamiltonian
\begin{equation}
	\hat H^{\rm E}_{\mathrm{int}}(t)=
	-\hat{\mathbf P}\cdot \mathbf E(t),
	\label{eq:Hint_EM}
\end{equation}	
with the spin-induced electric polarization~\cite{KNBMechanism,Cu2OSeO3Magnetoelectric} $\hat{\mathbf P}=\lambda_E\sum_{\langle i,j\rangle}\mathbf e_{ij}\times\left(\hat{\mathbf S}_i\times \hat{\mathbf S}_j\right)$. The detail study on the manipulation via magnetic field is given in Ref.~\cite{SupplementalMaterial}. Thus, the transition matrix elements between two eigenstates $|\psi_n\rangle$ and $|\psi_m\rangle$ are $P_{mn} = \langle \psi_m|\hat{\mathbf P}|\psi_n\rangle$.

For the periodic triangular cluster, take $\hat R_{\mathrm L}(60^\circ)$ to denote the counterclockwise lattice rotation that maps the site $i$ to $r(i)$.~Thus, for a local spin operator $\hat{\mathbf S}_i$, one has $\hat R_{\mathrm L}(60^\circ)\,\hat{\mathbf S}_i\,\hat R_{\mathrm L}^{-1}(60^\circ)=\hat{\mathbf S}_{r(i)}$.~Because the in-plane DMI vectors $\mathbf{D}_{ij}$ co-rotate with the bond geometry, this lattice rotation must be accompanied by a global spin rotation about the $z$ axis, i.e., $\hat R_{\mathrm S}(60^\circ)=\exp\!\left(-i\frac{\pi}{3}\sum_i \hat S_i^z\right)$, leading to the physical symmetry operation $\hat R_{\rm phys}=\hat R_{\mathrm L}(60^\circ)\hat R_{\mathrm S}(60^\circ)=\hat R_{\mathrm S}(60^\circ)\hat R_{\mathrm L}(60^\circ)$. That means, the periodic Hamiltonian is invariant, i.e., $[\hat H,\hat R_{\rm phys}]=0$. For a skyrmion with $N=19$ spin , however, a $2\pi$ spin rotation produces a minus sign, i.e., $\hat R_{\rm phys}^{\,6}=-\mathbb I$.~In the following numerical implementation, we therefore consider to use the gauge-equivalent operator $\hat R_{60^\circ}=e^{i\pi/6}\hat R_{\rm phys}$, which satisfies $\hat R_{60^\circ}^{\,6}=\mathbb I$. This operator redefinition has no influence on the commutation relation with the Hamiltonian, the operators transformation properties, or any transition matrix element, while allowing the eigenstates to be chosen as $\hat{R}_{60^\circ}|\psi_n\rangle = e^{i\pi \nu_n/3}|\psi_n\rangle$, where $\nu_n\in\{0,1,\ldots,5\}$ is the pseudo-angular-momentum quantum number defined modulo $6$. The symmetry of the transition operator~\cite{Wigner1931,Edmonds,BrinkSatchler}~for low-energy quantum-skyrmion states~\cite{QuantumSkyrmionFrustratedMagnet,Frustrated3}~then results in $\hat{ R}_{60^\circ} \hat{\mathbf P} (\hat{R}_{60^\circ})^{-1} =e^{i\pi \Delta \nu/3}\, \hat{\mathbf P}$, with the discrete angular-momentum transfer $\Delta \nu$ carried by the operator, and thus $\langle \psi_m|\hat{\mathbf P}|\psi_n\rangle=e^{i\frac{\pi}{3}(\nu_n-\nu_m+\Delta \nu)} \langle \psi_m|\hat{\mathbf P}|\psi_n\rangle$. Hence, nonzero matrix element exists only if $\nu_m-\nu_n \equiv \Delta \nu \pmod 6$, which is the discrete $C_6$ counterpart of typical angular-momentum selection rule for electromagnetic transitions~\cite{Wigner1931,Edmonds,BrinkSatchler}.

\begin{figure}[tb]
	\centering
	\includegraphics[width=\columnwidth]{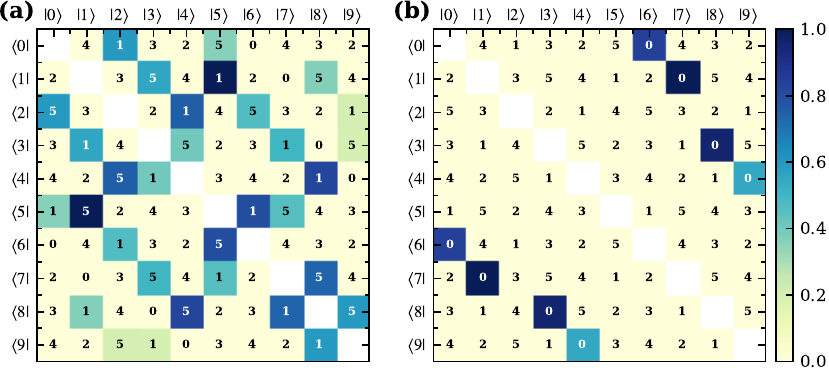}
	\caption{Electric-dipole transition matrix elements of the low-energy quantum-skyrmion eigenstates at $J=-0.5$, $B=0.45$, and $\lambda=0.65$. Panels (a) and (b) show the electric-dipole transition matrix elements for in-plane and out-of-plane electric fields, quantified by $P_{mn}^{\perp}$ and $P_{mn}^{z}$, respectively. For each coupling type, we normalize all transition matrix elements by their own largest off-diagonal matrix element, with the diagonal matrix elements omitted. Each off-diagonal cell for the matrix element $\langle m|\hat P|n\rangle$ is labeled by $(\nu_n-\nu_m)\bmod 6$.}
	\label{fig:selection_rules}
\end{figure}

To identify allowed electric transitions explicitly, we introduce circular transverse component
\begin{equation}
	\hat P_{\pm}=
	\hat P_x\pm i\hat P_y,
	\label{eq:circular_components_main}
\end{equation}
together with the longitudinal component $\hat P_z$. The circular transverse components transform as $\pm1$ tensor operators under rotations about the $z$ axis, so it carries $\Delta \nu=\pm1$ and thus connects only the states of sectors differing by one unit modulo $6$~\cite{Edmonds,BrinkSatchler}. By contrast, the longitudinal component is invariant under the joint rotation so it carries $\Delta \nu=0$ and thus couples only the states within the same pseudo-angular-momentum sector. These symmetry rules reflect the broader principle: in topological or symmetry-structured quantum systems, coherent control is governed by operator-resolved transitions rather than by spectral proximity alone~\cite{SkyrmionArray2}.

We now study the transition selection rules when the PBC is broken. By taking the parameters $J=-0.5$, $B=0.45$, and $\lambda=0.65$ where the boundary interpolation breaks the global $C_6$ symmetry while preserves the scalar chirality $Q_n \simeq 0.5$, we show the corresponding transition matrix elements $P_{mn}^{z}\equiv|\langle m|P_z|n\rangle|$ and $P_{mn}^{\perp} \equiv \sqrt{|\langle m|P_x|n\rangle|^2+ |\langle m|P_y|n\rangle|^2}$  in Figs.~\ref{fig:selection_rules}(a)~and~\ref{fig:selection_rules}(b), respectively. Remarkably, $P_\pm$ preserves the corresponding off-diagonal pattern, while $P_z$ retains a predominantly near-block-diagonal structure, indicating that the $C_6$ selection rules remain approximately valid despite substantial boundary perturbations. This persistence reflects the predominantly rotational character of the electric-polarization operator. Since the DMI Hamiltonian breaks inversion symmetry, the parity does not provide an additional good quantum number. For comparisons, we also study an inversion-symmetric frustrated-exchange model, whose eigenstates can be classified by parity and thus the electromagnetic transition pattern becomes more restrictive~\cite{SupplementalMaterial}.

\textcolor{blue}{\it Single phonon control of quantum skyrmions.---} In contrast to the microwave photons, the SAW phonons have the small mode volume, they are easy to reach the strong coupling with the quantum skyrmion. Thus, we now study the single-phonon control on  the two lowest energy levels of  the quantum skyrmion (hereafter, called as the skyrmion qubit). To maintain a material-grounded estimate, we use parameters~\cite{JanusNanoscale2023}: the DMI strength $D=1.78~\mathrm{meV}$ and the exchange interaction $J=-0.89~\mathrm{meV}$, the external magnetic field is chosen as $B_{\mathrm{ext}}=14.0~\mathrm{T}$. We adopt $\lambda_E=2\times 10^{-30}\,\mathrm{C\cdot m}$ for the spin-driven electric polarization~\cite{JanusNanoscale2023,KNBMechanism}. With these parameters, the exact-diagonalization spectrum of the $N=19$ cluster yields a transition frequency $\omega_{01}/2\pi=4.54\,\mathrm{GHz}$ between two energy-lowest states $|0\rangle$ and $|1\rangle$, with the near transitions $\omega_{02}/2\pi=4.79\,\mathrm{GHz}$ and $\omega_{12}/2\pi=0.25\,\mathrm{GHz}$ yielding a large anharmonicity~\cite{SkyrmionQubitFirstReview}. More importantly, the electromagnetic transition selection rules permit the transition~$|0\rangle\leftrightarrow|1\rangle$, while forbidding the transitions~$|0\rangle\leftrightarrow|2\rangle$ and $|1\rangle\leftrightarrow|2\rangle$~\cite{SupplementalMaterial}. Therefore, the two energy-lowest skyrmionic states form a well-isolated effective two-level system: a resonant external drive can coherently manipulate the $|0\rangle\leftrightarrow|1\rangle$ transition without leading to leakage out from computational subspace. This anharmonicity provides a spectral separation from higher levels, further reinforcing the validity of the qubit approximation.

Once restricted to the two states of the skyrmion qubit, the hybrid system is described by the Hamiltonian~\cite{JaynesCummings1963}
\begin{equation}
	\hat H_{\mathrm{JC}}=
	\omega_m a^\dagger a+
	\frac{\omega_{01}}{2}\sigma_z+
	g_m\left(a^\dagger \sigma_- + a\sigma_+\right),
	\label{eq:HJC_main}
\end{equation}
by replacing the classical electric field $E(t)$ in Eq.~(\ref{eq:Hint_EM}) with a quantized one generated by quantized single-mode SAW through piezoelectric effect~\cite{QuantumAcousticsTransducer1}, described by  the annihilation $a$ and creation operator $a^\dagger$, respectively. The coupling strength $g_m$ between the skyrmion qubit and a single-mode SAW is $g_m= E_{\mathrm{zp}}\,\bigl| \langle 1|\hat{\mathbf P}\!\cdot\!\mathbf e_m|0\rangle\bigr|$ by projecting the spin-induced polarization operator $\hat{\mathbf P}$ in Eq.~(\ref{eq:Hint_EM}) onto the skyrmion qubit subspace, with the polarization direction $\mathbf e_m$ of SAW-induced electric field.  If we consider that a baseline SAW mode resonates with the transition $|0\rangle\leftrightarrow|1\rangle$, i.e., $f_{\mathrm{base}}/2\pi=\omega_{01}/2\pi=4.54\,\mathrm{GHz}$,  then the zero-point electric field  induced by the piezoelectric effect is $E_{\mathrm{zp}} \simeq 1.82 \times 10^2\,\mathrm{V/m}$~\cite{QuantumAcousticsTransducer1,HybridSAWSkyrmion}. This yields the coupling strength between the skyrmion qubit and a SAW phonon
\begin{equation}
	g_m/2\pi \simeq 0.74\,\mathrm{MHz},
	\label{eq:gm_numeric}
\end{equation}
for skyrmion with $N=19$ spins, which is already close to the largest size accessible within our available exact-diagonalization resources. We mention that Eq.~(\ref{eq:HJC_main}) becomes classical manipulation on the skyrmion qubit when the quantized SAW is replaced by classical one.

To assess whether this smallest numerically accessible skyrmion can reach the strong-coupling regime, we estimate the intrinsic dissipation of the skyrmion qubit by regarding it as a magnon-like excitation and parameterizing its linewidth by a Gilbert damping factor $\alpha_G$, which is the standard phenomenological measure of magnetic energy loss in Landau-Lifshitz-Gilbert dynamics~\cite{GilbertDamping-Permalloy} and can be small in two-dimensional van der Waals ferromagnets~\cite{UltralowGilbert2DVdW2025}. That is, we take $\alpha_G=3\times10^{-5}$, corresponding to the decay $\gamma_{sk}/2\pi \sim \alpha_G \omega_{01}/2\pi \simeq 0.136\,\mathrm{MHz}.$ Comparing this linewidth with the single-phonon coupling strength in Eq.~(\ref{eq:gm_numeric}), we find $g_m/\gamma_{sk} \sim 5.44,$ showing that the strong-coupling can be reached even in the minimal cluster $N=19$ under above parameters.

To reveal the skyrmion size-dependent $g_m$, we consider a smooth skyrmion texture in a continuum limit~\cite{SupplementalMaterial}, where the electric polarization $\mathbf P(\mathbf r)$  in Eq.~(\ref{eq:Hint_EM}) is replaced by $\mathbf P_{\mathrm{tot}}=\int d^2\mathbf r\,\mathbf P(\mathbf r)$. For inverse-DMI magnetoelectricity, the local polarization density has the generic form $\mathbf P(\mathbf r)\propto(\mathbf m\cdot\nabla)\mathbf m-\mathbf m(\nabla\cdot\mathbf m)$~\cite{KNBMechanism,Cu2OSeO3Magnetoelectric}, with the magnetization $\mathbf m$.~Upon scaling the skyrmion radius as $R$ while preserving its dimensionless profile, the area element contributes $R^2$ and the gradient contributes $R^{-1}$, yielding the polarization scales linearly with size, i.e., $\mathbf P_{\mathrm{tot}}\propto R$. That means, the single-phonon coupling strength scales directly with the skyrmion size, i.e., $g_m \propto R$. For the parameters used above, the continuum exchange-DMI length scale gives $R_{\mathrm{cl}}\simeq3.14a$, compared with the effective radius $R_{N=19}\simeq2a$ of the minimal $N=19$ cluster, i.e., $R_{\mathrm{cl}}/R_{N=19} \simeq 1.57$. Thus, the $N=19$ exact-diagonalization result should be regarded as a conservative lower-bound estimate, with larger skyrmions offering enhanced spin-acoustic coupling. Extrapolating $g_m \propto R$, a coupling strength of $10\,\mathrm{MHz}$ corresponds to $R\simeq27a$, involving about $10^3$ spins. Although the diagonalization for such sizes remains challenge due to the exponential Hilbert-space growth, the continuum scaling quantitatively indicates that larger skyrmions enable the strong-coupling between the skyrmion qubit and a single-mode SAW phonon.

\textcolor{blue}{\textit{Conclusion.---}}
In summary, we study coherent manipulation on a quantum skyrmion. Exact diagonalization of a finite Dzyaloshinskii-Moriya cluster identifies a low-energy skyrmionic states that are non-degenerate and preserve their scalar chirality even under coherent superposition, providing a natural logical subspace for skyrmion qubit. Within these states, the lattice-spin symmetry imposes electromagnetic transition selection rules, thereby enabling polarization-selective addressing of skyrmionic states. By coupling the symmetry-allowed electric-dipole transition with the electric field induced by a quantized SAW via piezoelectric effect, we show that the skyrmion-phonon interaction can enter the strong-coupling regime. The minimal quantum cluster considered here provides a conservative microscopic benchmark rather than an ultimate limit on the achievable skyrmion-phonon coupling. Larger skyrmionic textures can further enhance the coupling through the increased spin-induced electric polarization with the skyrmion size. Our results establish a microscopic connection between many-body spin physics and quantum acoustodynamics, and enable quantum skyrmions as compact information storage elements for hybrid on-chip quantum information processing.

{\it Acknowledgements.}
This work was supported by the National Natural Science Foundation of China Grants No. 12374483, No. 92365209, and No. 62474012.
	
\bibliography{QuantumSyrmion_refs}

@article{PhysRevLett.132.193601,
  title = {Magnon-Skyrmion Hybrid Quantum Systems: Tailoring Interactions via Magnons},
  author = {Pan, Xue-Feng and Li, Peng-Bo and Hei, Xin-Lei and Zhang, Xichao and Mochizuki, Masahito and Li, Fu-Li and Nori, Franco},
  journal = {Phys. Rev. Lett.},
  volume = {132},
  issue = {19},
  pages = {193601},
  numpages = {10},
  year = {2024},
  month = {May},
  publisher = {American Physical Society},
}

@article{SharmaPsaroudaki2026,
  author = {Sharma, Sanchar and Psaroudaki, Christina},
  title = {Optical Signatures of Quantum Skyrmions},
  journal = {Phys. Rev. Lett.},
  volume = {136},
  pages = {016701},
  year = {2026},
}

@article{HallerDiazBelzigSchmidt2024,
  author = {Haller, Andreas and D{\'i}az, Sebasti{\'a}n A. and Belzig, Wolfgang and Schmidt, Thomas L.},
  title = {Quantum Magnetic Skyrmion Operator},
  journal = {Phys. Rev. Lett.},
  volume = {133},
  pages = {216702},
  year = {2024},
}

@article{Oleg,
  title = {Quantum Regime of a Two-Dimensional Phonon Cavity},
  author = {Bolgar, Aleksey N. and Zotova, Julia I. and Kirichenko, Daniil D. and Besedin, Ilia S. and Semenov, Aleksander V. and Shaikhaidarov, Rais S. and Astafiev, Oleg V.},
  journal = {Phys. Rev. Lett.},
  volume = {120},
  issue = {22},
  pages = {223603},
  numpages = {5},
  year = {2018},
  month = {May},
  publisher = {American Physical Society},
}

@article{SupplementalMaterial,
  author  = {{Supplemental Material}},
  title   = {Supplemental Material for ``Coherent Manipulation of Quantum Skyrmion via Surface Acoustic Wave Phonons''},
  journal = {Phys. Rev. Lett.},
  note    = {See Supplemental Material at [URL will be inserted by publisher] for the visualization of low-energy DMI textures, the quantum-to-classical regression analysis, parity-resolved selection rules, and the classical DMI-skyrmion-radius estimate.}
}

@article{Heisenberg1928,
  author  = {Heisenberg, W.},
  title   = {Zur Theorie des Ferromagnetismus},
  journal = {Z. Phys.},
  volume  = {49},
  number  = {9--10},
  pages   = {619--636},
  year    = {1928},
}

@article{Dzyaloshinskii1958,
  author  = {Dzyaloshinskii, I. E.},
  title   = {A Thermodynamic Theory of ``Weak'' Ferromagnetism of Antiferromagnetics},
  journal = {J. Phys. Chem. Solids},
  volume  = {4},
  number  = {4},
  pages   = {241--255},
  year    = {1958},
}

@article{Moriya1960,
  author  = {Moriya, T.},
  title   = {Anisotropic Superexchange Interaction and Weak Ferromagnetism},
  journal = {Physical Review},
  volume  = {120},
  number  = {1},
  pages   = {91--98},
  year    = {1960},
}

@article{SkyrmionReview2,
  author  = {Petrovic, A. P. and Psaroudaki, C. and Fischer, P. and
             Garst, M. and Panagopoulos, C.},
  title   = {Quantum Properties and Functionalities of Magnetic Skyrmions},
  journal = {Rev. Mod. Phys.},
  volume  = {97},
  pages   = {031001},
  year    = {2025},
}

@article{Chen2025PiezomagneticSAW,
  author  = {Chen, Yu-Yuan and Wang, Jia-Heng and Song, Lu Ning and Liu, Yu-Xi},
  title   = {Manipulation of Magnetic Systems by Quantized Surface Acoustic
             Waves via the Piezomagnetic Effect},
  journal = {Phys. Rev. Appl.},
  volume  = {23},
  pages   = {034013},
  year    = {2025},
}

@article{JaynesCummings1963,
  author  = {Jaynes, E. T. and Cummings, F. W.},
  title   = {Comparison of Quantum and Semiclassical Radiation Theories with Application to the Beam Maser},
  journal = {Proc. IEEE},
  volume  = {51},
  number  = {1},
  pages   = {89--109},
  year    = {1963},
}

@article{satzinger2018quantum,
  author  = {Satzinger, Kevin J. and Zhong, Y. P. and Chang, H.-S. and Peairs, Gregory A. and Bienfait, A. and Chou, Ming-Han and Cleland, A. Y. and Conner, C. R. and Dumur, \'E. and Grebel, J. and Gutierrez, I. and November, B. H. and Povey, R. G. and Whiteley, S. J. and Awschalom, D. D. and Schuster, D. I. and Cleland, A. N.},
  title   = {Quantum control of surface acoustic wave phonons},
  journal = {Nature},
  volume  = {563},
  pages   = {661--665},
  year    = {2018},
}

@article{SkyrmionQubitPhonon2024,
  title = {Skyrmion-mechanical hybrid quantum systems: Manipulation of skyrmion qubits via phonons},
  author = {Pan, Xue-Feng and Hei, Xin-Lei and Yao, Xiao-Yu and Chen, Jia-Qiang and Ren, Yu-Meng and Dong, Xing-Liang and Qiao, Yi-Fan and Li, Peng-Bo},
  journal = {Phys. Rev. Research},
  volume = {6},
  issue = {2},
  pages = {023067},
  numpages = {16},
  year = {2024},
  month = {Apr},
  publisher = {American Physical Society},
}

@article{GilbertDamping-Permalloy,
  author  = {Liu, T. and Zhang, Y. and K\u{a}kol, Z. and others},
  title   = {Experimental Investigation of Temperature-Dependent Gilbert Damping in Permalloy Thin Films},
  journal = {Sci. Rep.},
  volume  = {6},
  pages   = {22890},
  year    = {2016},
}

@article{UltralowGilbert2DVdW2025,
  author  = {Chen, Weizhao and Zhang, Yu and Liu, Yi and Yuan, Zhe},
  title   = {Symmetry-Forbidden Intraband Transitions Leading to Ultralow Gilbert Damping in van der Waals Ferromagnets},
  journal = {Phys. Rev. Lett.},
  volume  = {135},
  number  = {17},
  pages   = {176704},
  year    = {2025},
}

@article{SuperconductingCircuit1,
  title = {Hybrid quantum circuits: Superconducting circuits interacting with other quantum systems},
  author = {Xiang, Ze-Liang and Ashhab, Sahel and You, J. Q. and Nori, Franco},
  journal = {Rev. Mod. Phys.},
  volume = {85},
  issue = {2},
  pages = {623--653},
  numpages = {0},
  year = {2013},
  month = {Apr},
  publisher = {American Physical Society},
}

@article{SuperconductingCircuit2,
  author  = {Gu, X. and Kockum, A. F. and Miranowicz, A. and Liu, Y.-X. and Nori, F.},
  title   = {Microwave photonics with superconducting quantum circuits},
  journal = {Phys. Rep.},
  volume  = {718-719},
  pages   = {1},
  year    = {2017}
}

@article{SuperconductingHybrid,
  author  = {Clerk, A. A. and Lehnert, K. W. and Bertet, P. and Petta, J. R. and Nakamura, Y.},
  title   = {Hybrid quantum systems with circuit quantum electrodynamics},
  journal = {Nat. Phys.},
  volume  = {16},
  pages   = {257},
  year    = {2020}
}

@article{SuperconductingHybridChip,
  author  = {Xu, X.-B. and Wang, W.-T. and Sun, L.-Y. and Zou, C.-L.},
  title   = {Hybrid superconducting photonic-phononic chip for quantum information processing},
  journal = {Chip},
  volume  = {1},
  pages   = {100016},
  year    = {2022}
}

@article{QuantumAcousticsSAWSuperconductingCoupling1,
  author  = {Gustafsson, M. V. and Aref, T. and Kockum, A. F. and Ekstr{\"o}m, M. K. and Johansson, G. and Delsing, P.},
  title   = {Propagating phonons coupled to an artificial atom},
  journal = {Science},
  volume  = {346},
  pages   = {207},
  year    = {2014}
}

@article{QuantumAcousticsTransducer1,
  author  = {Schuetz, M. J. A. and Kessler, E. M. and Giedke, G. and Vandersypen, L. M. K. and Lukin, M. D. and Cirac, J. I.},
  title   = {Universal quantum transducers based on surface acoustic waves},
  journal = {Phys. Rev. X},
  volume  = {3},
  pages   = {031031},
  year    = {2015}
}

@article{QuantumAcousticsArticle2,
  author  = {Manenti, R. and Kockum, A. F. and Patterson, A. and Behrle, T. and Rahamim, J. and Tancredi, G. and Nori, F. and Leek, P. J.},
  title   = {Circuit quantum acoustodynamics with surface acoustic waves},
  journal = {Nat. Commun.},
  volume  = {8},
  pages   = {975},
  year    = {2017}
}

@article{SAWEntanglement1,
  author  = {Bienfait, A. and Satzinger, K. J. and Zhong, Y. P. and Chang, H.-S. and Chou, M.-H. and Conner, C. R. and Dumur, {\'{E}}. and Grebel, J. and Peairs, G. A. and Povey, R. G. and Cleland, A. N.},
  title   = {Phonon-mediated quantum state transfer and remote qubit entanglement},
  journal = {Science},
  volume  = {364},
  pages   = {368},
  year    = {2019}
}

@article{QuantumAcousticsSAWSuperconductingCoupling2,
  author  = {Zeng, G.-H. and Zhang, Y. and Bolgar, A. N. and He, D. and Li, B. and Ruan, X.-H. and Zhou, L. and Kuang, L. M. and Astafiev, O. V. and Liu, Y.-X. and Peng, Z. H.},
  title   = {Quantum versus classical regime in circuit quantum acoustodynamics},
  journal = {New J. Phys.},
  volume  = {23},
  pages   = {123001},
  year    = {2021}
}

@article{SAWEntanglement2,
  author  = {Andersson, G. and Jolin, S. W. and Scigliuzzo, M. and Borgani, R. and Thol{\'e}n, M. O. and Hern{\'a}ndez, J. C. R. and Shumeiko, V. and Haviland, D. B. and Delsing, P.},
  title   = {Squeezing and multimode entanglement of surface acoustic wave phonons},
  journal = {PRX Quantum},
  volume  = {3},
  pages   = {010312},
  year    = {2022}
}

@book{SAWBackgroundBook1,
  author    = {Datta, S.},
  title     = {Surface Acoustic Wave Devices},
  publisher = {Prentice-Hall},
  address   = {Upper Saddle River, NJ},
  year      = {1986}
}

@book{SAWBackgroundBook2,
  author    = {Morgan, D.},
  title     = {Surface Acoustic Wave Filters},
  publisher = {Academic Press},
  address   = {Boston},
  year      = {2007}
}

@article{SAWCavity,
  author  = {Haydl, W. H. and Dischler, B. and Hiesinger, P.},
  title   = {Multimode SAW resonators: a method to study the optimum resonator design},
  journal = {Proc. IEEE Ultrason. Symp.},
  volume  = {1},
  pages   = {287},
  year    = {1976}
}

@article{QuantumAcousticsMultimodeSAW,
  author  = {Moores, B. A. and Sletten, L. R. and Viennot, J. J. and Lehnert, K. W.},
  title   = {Cavity quantum acoustic device in the multimode strong coupling regime},
  journal = {Phys. Rev. Lett.},
  volume  = {120},
  pages   = {227701},
  year    = {2018}
}

@article{SAWCavityMultimodeTunableCoupling,
  author  = {Ruan, X. H. and Li, L. and Liang, G. and Zhao, S. and Wang, J. and Bu, Y. and Chen, B. and Song, X. and Li, X. and Zhang, H. and Wang, J. and Zhao, Q. and Xu, K. and Fan, H. and Liu, Y. X. and Zhang, J. and Peng, Z. H. and Xiang, Z. C. and Zheng, D. N.},
  title   = {Tunable coupling of a quantum phononic resonator to a transmon qubit via galvanic-contact flip-chip architecture},
  journal = {Appl. Phys. Lett.},
  volume  = {125},
  pages   = {052603},
  year    = {2024}
}

@article{SkyrmionFirstTheory,
  author  = {Bogdanov, A. N. and R{\"o}{$\beta$}ler, U. K.},
  title   = {Chiral symmetry breaking in magnetic thin films and multilayers},
  journal = {Phys. Rev. Lett.},
  volume  = {87},
  pages   = {037203},
  year    = {2001}
}

@article{SkyrmionReviewDevices1,
  author  = {Finocchio, G. and B{\"u}ttner, F. and Tomasello, R. and Carpentieri, M. and Kl{\"a}ui, M.},
  title   = {Magnetic skyrmions: from fundamental to applications},
  journal = {J. Phys. D: Appl. Phys.},
  volume  = {49},
  pages   = {423001},
  year    = {2016}
}

@article{SkyrmionReviewDevices2,
  author  = {Fert, A. and Reyren, N. and Cros, V.},
  title   = {Magnetic skyrmions: advances in physics and potential applications},
  journal = {Nat. Rev. Mater.},
  volume  = {2},
  pages   = {17031},
  year    = {2017}
}

@article{SkyrmionReviewDevices3,
  author  = {Zhang, X. and Zhou, Y. and Song, K. M. and Park, T. E. and Xia, J. and Ezawa, M. and Liu, X. and Zhao, W. and Zhao, G. and Woo, S.},
  title   = {Skyrmion-electronics: writing, deleting, reading and processing magnetic skyrmions toward spintronic applications},
  journal = {J. Phys.: Condens. Matter},
  volume  = {32},
  pages   = {143001},
  year    = {2020}
}

@article{SkyrmionReview1,
  author  = {Reichhardt, C. and Reichhardt, C. J. O. and Milo{\v{s}}evi{\'c}, M. V.},
  title   = {Statics and dynamics of skyrmions interacting with disorder and nanostructures},
  journal = {Rev. Mod. Phys.},
  volume  = {94},
  pages   = {035005},
  year    = {2022}
}

@article{SkyrmionQubitFirstTheory,
  author  = {Psaroudaki, C. and Panagopoulos, C.},
  title   = {Skyrmion qubits: a new class of quantum logic elements based on nanoscale magnetization},
  journal = {Phys. Rev. Lett.},
  volume  = {127},
  pages   = {067201},
  year    = {2021}
}

@article{SkyrmionQubitFirstTheoryHelicity,
  author  = {Psaroudaki, C. and Panagopoulos, C.},
  title   = {Skyrmion helicity: Quantization and quantum tunneling effects},
  journal = {Phys. Rev. B},
  volume  = {106},
  pages   = {104422},
  year    = {2022}
}

@article{SkyrmionQubitFirstReview,
  author  = {Psaroudaki, C. and Peraticos, E. and Panagopoulos, C.},
  title   = {Skyrmion qubits: challenges for future quantum computing applications},
  journal = {Appl. Phys. Lett.},
  volume  = {123},
  pages   = {260501},
  year    = {2023}
}

@article{SkyrmionQubitQuantumComputing,
  author  = {Xia, J. and Zhang, X. C. and Liu, X. X. and Zhou, Y. and Ezawa, M.},
  title   = {Universal quantum computation based on nanoscale skyrmion helicity qubits in frustrated magnets},
  journal = {Phys. Rev. Lett.},
  volume  = {130},
  pages   = {106701},
  year    = {2023}
}

@article{QuantumSkyrmionFrustratedMagnet,
  author  = {Lohani, S. and Hickey, C. and Masell, J. and Rosch, A.},
  title   = {Quantum skyrmions in frustrated ferromagnets},
  journal = {Phys. Rev. X},
  volume  = {9},
  pages   = {041063},
  year    = {2019}
}

@article{QuantumSkyrmionStability,
  author  = {Salvati, F. and Katsnelson, M. I. and Bagrov, A. A. and Westerhout, T.},
  title   = {Stability of a quantum skyrmion: Projective measurements and the quantum Zeno effect},
  journal = {Phys. Rev. B},
  volume  = {109},
  pages   = {064409},
  year    = {2024}
}

@article{QuantumSkyrmionPatterns,
  author  = {Mazurenko, V. V. and Iakovlev, I. A. and Sotnikov, O. M. and Katsnelson, M. I.},
  title   = {Estimating Patterns of Classical and Quantum Skyrmion States},
  journal = {J. Phys. Soc. Jpn.},
  volume  = {92},
  pages   = {081004},
  year    = {2023}
}

@article{QuantumSkyrmionControlledCreation,
  author  = {Siegl, P. and Vedmedenko, E. Y. and Stier, M. and Thorwart, M. and Posske, T.},
  title   = {Controlled creation of quantum skyrmions},
  journal = {Phys. Rev. Research},
  volume  = {4},
  pages   = {023111},
  year    = {2022}
}

@article{BoundaryTuningSkyrmion,
  author  = {Spethmann, J. and Vedmedenko, E. Y. and Wiesendanger, R. and Kubetzka, A. and von Bergmann, K.},
  title   = {Zero-field skyrmionic states and in-field edge-skyrmions induced by boundary tuning},
  journal = {Commun. Phys.},
  volume  = {5},
  pages   = {19},
  year    = {2022}
}

@article{SkyrmionicQubitsDMI,
  author  = {Sticlet, D. and Tetean, R. and Tiusan, C.},
  title   = {Skyrmionic qubits stabilized by Dzyaloshinskii-Moriya interaction as platforms for qubits and quantum gates},
  journal = {Phys. Rev. B},
  volume  = {112},
  pages   = {195435},
  year    = {2025}
}

@article{HybridSAWSkyrmion,
  author  = {Chen, Y.-Y. and Peng, Z. and Liu, Y.-X.},
  title   = {Hybrid quantum surface acoustic wave with skyrmion qubit for quantum information processing},
  journal = {Phys. Rev. Lett.},
  volume  = {136},
  pages   = {013801},
  year    = {2026}
}

@article{KNBMechanism,
  author  = {Katsura, H. and Nagaosa, N. and Balents, A. V.},
  title   = {Spin current and magnetoelectric effect in noncollinear magnets},
  journal = {Phys. Rev. Lett.},
  volume  = {95},
  pages   = {057205},
  year    = {2005}
}

@article{Cu2OSeO3Magnetoelectric,
  author  = {Seki, S. and Yu, X. Z. and Ishiwata, S. and Tokura, Y.},
  title   = {Observation of magnetic skyrmions in a multiferroic material},
  journal = {Science},
  volume  = {336},
  pages   = {198},
  year    = {2012}
}

@article{JanusNanoscale2023,
  author = {Han, Yue-tong and Ji, Wei-xiao and Wang, Pei-Ji and Li, Ping and Zhang, Chang-Wen},
  title = {Strain-tunable skyrmions in two-dimensional monolayer Janus magnets},
  journal = {Nanoscale},
  volume = {15},
  number = {14},
  pages = {6830-6837},
  year = {2023},
  month = {04},
}

@article{SkyrmionMotionControl1,
  author  = {Woo, S. and Litzius, K. and Kr{\"u}ger, B. and Im, M. Y. and Caretta, L. and Richter, K. and Mann, M. and Krone, A. and Reeve, R. M. and Weigand, M. and Agrawal, P. and Lemesh, I. and Mawass, M. A. and Fischer, P. and Kl{\"a}ui, M. and Beach, G. S. D.},
  title   = {Observation of room-temperature magnetic skyrmions and their current-driven dynamics in ultrathin metallic ferromagnets},
  journal = {Nat. Mater.},
  volume  = {15},
  pages   = {501},
  year    = {2016}
}

@article{SkyrmionMotionControl2,
  author  = {Zhang, S. L. and Wang, W. W. and Burn, D. M. and Peng, H. and Berger, H. and Bauer, A. and Pfleiderer, C. and van der Laan, G. and Hesjedal, T.},
  title   = {Manipulation of skyrmion motion by magnetic field gradients},
  journal = {Nat. Commun.},
  volume  = {9},
  pages   = {2115},
  year    = {2018}
}

@article{SkyrmionMotionControl3,
  author  = {Yokouchi, T. and Sugimoto, S. and Rana, B. and Seki, S. and Ogawa, N. and Kasai, S. and Otani, Y.},
  title   = {Creation of magnetic skyrmions by surface acoustic waves},
  journal = {Nat. Nanotechnol.},
  volume  = {15},
  pages   = {361},
  year    = {2020}
}

@article{SkyrmionModulationField,
  author  = {Wang, W. W. and Beg, M. and Zhang, B. and Kuch, W. and Fangohr, H.},
  title   = {Driving magnetic skyrmions with microwave fields},
  journal = {Phys. Rev. B},
  volume  = {92},
  pages   = {020403(R)},
  year    = {2015}
}

@article{SkyrmionArray2,
  author  = {Tengdin, P. and Truc, B. and Sapozhnik, A. and Kong, L. and del Ser, N. and Gargiulo, S. and Madan, I. and Sch{\"o}nenberger, T. and Baral, P. R. and Che, P. and Magrez, A. and Grundler, D. and R{\o}nnow, H. M. and Lagrange, T. and Zang, J. D. and Rosch, A. and Carbone, F.},
  title   = {Imaging the ultrafast coherent control of a skyrmion crystal},
  journal = {Phys. Rev. X},
  volume  = {12},
  pages   = {041030},
  year    = {2022}
}

@book{Wigner1931,
  author    = {Wigner, E.},
  title     = {Group Theory and Its Application to the Quantum Mechanics of Atomic Spectra},
  publisher = {Academic},
  address   = {New York},
  year      = {1931}
}

@book{Edmonds,
  author    = {Edmonds, A. R.},
  title     = {Angular Momentum in Quantum Mechanics},
  publisher = {Princeton University Press},
  address   = {Princeton},
  year      = {1957}
}

@book{BrinkSatchler,
  author    = {Brink, D. M. and Satchler, G. R.},
  title     = {Angular Momentum},
  publisher = {Clarendon Press},
  address   = {Oxford},
  year      = {1968}
}

@article{Gustafsson2014,
  author  = {Gustafsson, Martin V. and Aref, Thomas and Kockum, Anton Frisk and Ekstr{\"o}m, Maria K. and Johansson, G{\"o}ran and Delsing, Per},
  title   = {Propagating phonons coupled to an artificial atom},
  journal = {Science},
  volume  = {346},
  number  = {6206},
  pages   = {207--211},
  year    = {2014}
}

@article{Manenti2017,
  author  = {Manenti, Riccardo and Kockum, Anton F. and Patterson, Andrew and Behrle, Tanja and Rahamim, Joseph and Tancredi, Giovanna and Nori, Franco and Leek, Peter J.},
  title   = {Circuit quantum acoustodynamics with surface acoustic waves},
  journal = {Nat. Commun.},
  volume  = {8},
  number  = {1},
  pages   = {975},
  year    = {2017}
}

@article{Frustrated3,
  author  = {Lohani, S. and Hickey, C. and Masell, J. and Egger, A.},
  title   = {Quantum skyrmions on the triangular lattice},
  journal = {Phys. Rev. B},
  volume  = {101},
  pages   = {024430},
  year    = {2020}
}
			
\end{document}